\documentclass[sigconf,10pt]{acmart}
\usepackage{physics}
\usepackage{amsmath}
\usepackage[utf8]{inputenc}
\usepackage{natbib}
\usepackage{graphicx}
\usepackage{subcaption}
\usepackage{pbox}
\usepackage{amsmath}
\usepackage{amsfonts}
\usepackage{flushend}
\usepackage{graphicx}
\usepackage{tabularx}
\usepackage{tabularx}
\usepackage{adjustbox}
\usepackage{algorithm}
\usepackage{algpseudocode}

\usepackage{booktabs} 
\usepackage{comment}
\usepackage{xcolor}
\usepackage{multirow}

\setcopyright{acmcopyright}
\copyrightyear{2021}
\acmYear{2021}
\acmDOI{10.1145/1122445.1122456}

\acmConference[SIGSPATIAL'21]{SIGSPATIAL '21: ACM 27th Annual International Conference On Mobile Computing And Networking}{October 25-29, 2021}{New Orleans, LA}
\acmBooktitle{New Orleans '21: ACM 27th Annual International Conference On Mobile Computing And Networking,
  October 25-29, 2021, New Orleans, LA}
\acmPrice{15.00}
\acmISBN{978-1-4503-XXXX-X/18/06}

\begin{document}

\title{Handling Device Heterogeneity for Deep Learning-based Localization} 


\author{Ahmed Shokry}
\affiliation{%
  \institution{Pennsylvania State University}
  \city{PA}
  \country{USA}}
\email{ahmed.shokry@psu.edu}

\author{Moustafa Youssef}
\affiliation{%
  \institution{American University in Cairo}
  \city{Cairo}
  \country{Egypt}}
\email{moustafa-youssef@aucegypt.edu}

\renewcommand{\shortauthors}{Shokry and Youssef}

\begin{abstract}
Deep learning-based fingerprinting is one of the current promising technologies for outdoor localization in cellular networks.
However, deploying such localization systems for heterogeneous phones affects their accuracy as the cellular received signal strength (RSS) readings vary for different types of phones.
In this paper, we introduce a number of techniques for addressing the phones heterogeneity problem in the deep-learning based localization systems. The basic idea is either to approximate a function that maps the cellular RSS measurements between different devices or to transfer the knowledge across them.

Evaluation of the proposed techniques using different Android phones on four independent testbeds shows that our techniques can improve the localization accuracy by more than 220\% for the four testbeds as compared to the state-of-the-art systems. This highlights the promise of the proposed device heterogeneity handling techniques for enabling a wide deployment of deep learning-based localization systems over different devices.
\end{abstract}



\maketitle
\section{Introduction}
Due to the high noise inherent in the wireless channel, fingerprinting based techniques cannot learn a good mapping between signals and locations. To address these limitations, a number of deep-learning based localization systems have been proposed recently \cite{deeploc,wang2017resloc}; which can can capture a more accurate mapping between signals and locations. However, when the training device(s) that are used to build the radio map and the test devices that are used for positioning during actual system operation are different, the system accuracy severely degrades. Hence, deploying such systems on different phone types is not a straightforward task as the RSS readings vary for the different kinds of phones, even at the same location and time. This problem is known as the
device heterogeneity problem \cite{park2011implications,shokry2017tale, shokry2020dynamicslam}.

The straightforward solution for the heterogeneity problem is to build and train different deep models using a fingerprint for each type of phones. Nonetheless, the range of available phone types in the market, which keeps growing every day, makes this process not scalable and have a high overhead.

Several techniques addressed this problem in classical localization~\cite{ibrahim2013enabling, park2011implications,figuera2011time, gu2021effect} and even in quantum localization~\cite{quantum_arx, device_indp_q, qhet, quantum_lcn, quantum_qce}.
These techniques attempt to obtain a mapping function that maps the RSS readings between the different types of phones \cite{ibrahim2013enabling, park2011implications, figuera2011time}. In \cite{ibrahim2013enabling}, the relative power technique was proposed. Its basic idea is to use the ratios of RSS readings instead of absolute values. Other techniques, e.g. \cite{park2011implications, figuera2011time}, try to learn a mapping between the RSS values of the different phone types using linear transformation with regression. To do that, the training device and each test device must both collect data at the same location and time. 
In \cite{tsui2009unsupervised}, an unsupervised learning approach is proposed. It determines the linear mapping function by learning from online measured RSSs. However, it requires a long learning period to get the accurate position. In \cite{della2010ad}, collaborative mapping between neighboring devices is used to determine the linear mapping function. However, it is required that at least one test device be the same as the training device.
Moreover, all the previous heterogeneity-handling techniques solve the heterogeneity problem only for the probabilistic models. Hence, new techniques are needed to handle the problem for deep models-based localization systems.

In this paper, we introduce a number of techniques for addressing the phone heterogeneity problem in deep-learning based localization systems. We first experiment with the traditional approaches for solving the device heterogeneity problem. Then, We propose two novel deep-learning based approaches that leverage the universal learning ability of deep models \cite{pan2009survey} to solve the heterogeneity problem. The first technique transfers the knowledge gained by training the deep model on a device to other devices by fine-tuning the deep learning model using a few samples for the tuning devices. The second technique models the heterogeneity problem as a Multi-Task Learning problem \cite{caruana1997multitask}. 
We have implemented and evaluated the different device heterogeneity solutions in the proposed framework on different Android phones in four different testbeds. Our results show that the proposed techniques can improve the localization accuracy by more than 220\% compared to the system without handling the heterogeneity. 
The rest of the paper is organized as follows: Section~\ref{method} presents the proposed general localization model we integrate our proposed heterogeneity handling techniques within. Section~\ref{hetsec} describes the details of the different heterogeneity handling techniques. We evaluate the different techniques in Section~\ref{evaluation}. Finally, we conclude the letter in Section~\ref{conclusion}.
\section{The localization model}
\label{method}

\begin{figure}[!tbp]
\centering
\includegraphics[width=6cm,height=4cm]{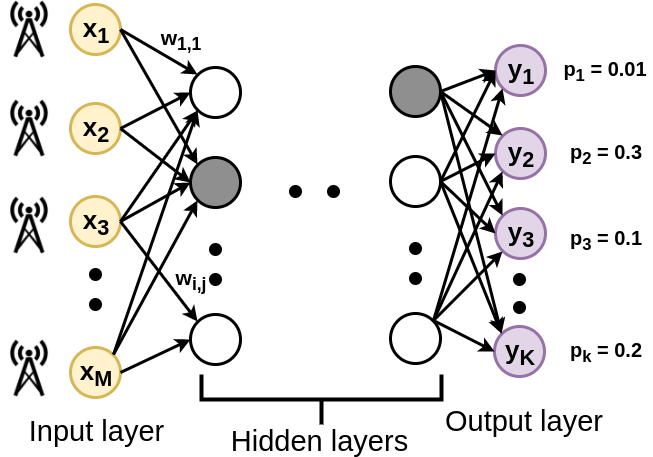}
     \caption{The proposed deep learning localization model. The input is the RSS coming from different cell towers in the environment (features). The output is a probability distribution over different reference locations in the area of interest. Gray-shaded neurons represent examples of nodes that have been temporary dropped-off to increase the model robustness and avoid over-training.}
     \label{fig:model}
\end{figure}

Figure~\ref{fig:model} shows the localization model proposed in this study. 
The input to the model is the RSS measurements collected by the phones (typically heterogeneous) and the output is a probability distribution for different reference locations in the area of interest. The location with the highest probability can be used as the estimated location or the center of mass of the different reference locations. 

To collect the training data for the reference locations, the cell phone scans for the cell towers at the reference locations (total $K$). According to the standard \cite{ibrahim2012cellsense}, up to seven towers of the total $M$ available cell towers in the area can be heard at any scan. For each heard cell tower, a pair (CID, RSS) is recorded, where CID represents the cell tower unique ID and RSS is the received signal strength from that tower in dBm. Different RSS scans are performed at each reference location to collect the training data.
The feature vector per sample is constructed from all cell towers detectable in the area of interest. Since not all these towers can be heard in every scan, the system assigns
zero to the towers that are not heard in a given scan. This allows us to have a fixed feature vector size. 
\section{Heterogeneity Solutions}
\label{hetsec}
In this section, we describe the different techniques to handle the devices heterogeneity problem.
The proposed techniques are grouped into two categories. The first is the traditional techniques which have been used before with the localization probabilistic models to solve the devices heterogeneity problem. The second is the deep learning-based techniques. 

\subsection{Traditional techniques}
These techniques try to approximate a function that maps the RSS vectors across the different heterogeneous devices.\\

\subsubsection{Linear transformation}
This technique assumes that the relation between the RSS coming from a cell tower at the different phones is linear. Hence, it attempts to approximate a line parameters for each cell-tower that maps the scalar RSS reading from one phone to another. To approximate this function, we collect RSS points as pairs of $(x_i,x_j)$ where $x_i$ and $x_j$ are the received signal strengths coming from a cell tower for devices $i$ and $j$, respectively, at the same location. Finally, we estimate parameters of the line that best fits those points using the least square method.
Once this mapping is obtained, the deep model is trained with the RSS values coming from a single device ($i$). Then, given a RSS vector coming from another device ($j$) and the learned line parameters for each cell tower, one can map the input RSS vector from device $j$ to device ($i$), which is then fed to the model to estimate the phone location.\\

\subsubsection{Power ratio}
This technique assumes that the ratio between the RSS values coming from different cell towers remains the same on the different phones. Hence, it transforms the RSS values (i.e. power) to a relative power. That is, instead of using the raw RSS from individual cell towers as its input, it uses the ratio of the cellular RSS readings from \textit{\textbf{each pair}} of cell towers. 
More formally, given a RSS features vector $X$= $(f_1,f_2,..,f_M)$ from $M$ cell towers in the environment, where $f_i$ is the RSS coming from cell tower $i$, this technique transforms $X$ to $X_{r}$= $(r_{1,2},r_{1,3},..,r_{M-1,M})$, where $r_{i,j} = \frac{f_i}{f_j}$ and $\vert X_{r} \vert = \dbinom {M}{2}$. The deep model is then trained and tested using the new feature vectors $X_{r}$ instead of $X$. \\

\subsubsection{Power difference}
This approach is similar to the previous approach but takes the power difference instead of the power ratio. It assumes that the difference between the RSS values coming from different cell towers remains the same on the different phones.

More formally, given a RSS features vector $X$= $(f_1,f_2,..,f_M)$ from $n$ cell-towers in the environment, where $f_i$ is the RSS coming from cell tower $i$, this technique transforms $X$ to $X_{d}$= $(d_{1,2},d_{1,3},..,d_{M-1,M})$, where $d_{i,j} = f_i - f_j$ and $\vert X_{d} \vert = \dbinom {M}{2}$.

Finally the deep model is trained and tested using the new feature vectors $X_{d}$ instead of $X$. \\

\begin{figure}[!t]
\centering
\includegraphics[width=0.4\textwidth]{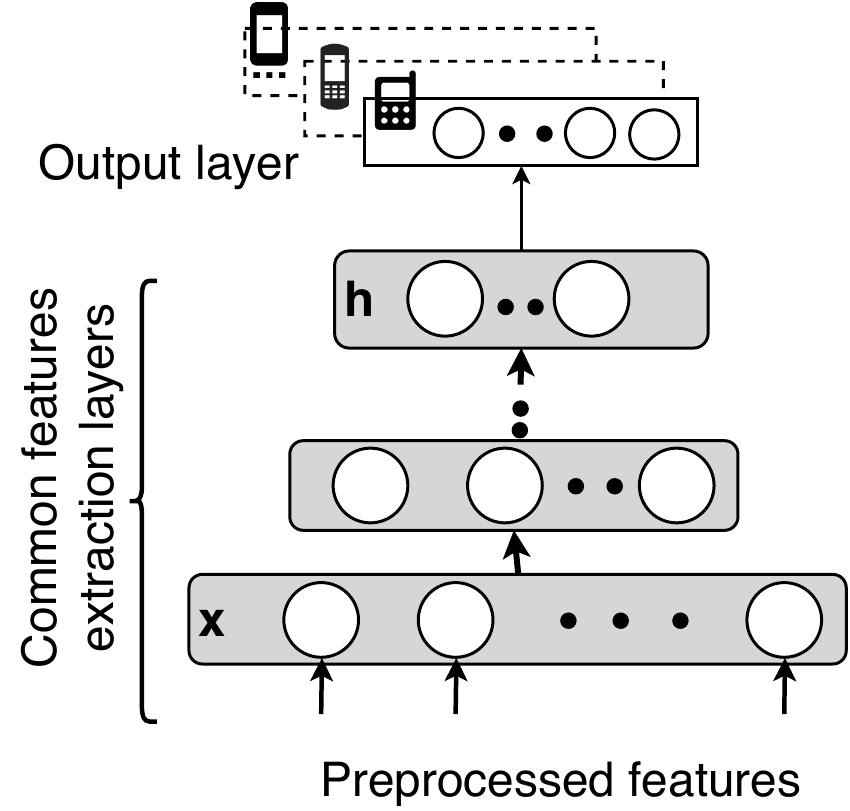}
     \caption{Transfer learning deep model structure. Input is the cell towers RSS information. Gray layers are responsible for extracting the relation between the RSS input features. The output layer is trained once as a part of the network using a specific device (solid layer) with a large number of samples. Then, the model weights are tuned using a small samples from the different devices (dashed layers).}
     \label{fig:network-tl}
\end{figure}
\subsection{Deep learning-based techniques}
Unlike the previous category, this one assumes that the relation between the RSS values coming from the different cell towers is not simple
. Hence, it employs a deep learning approaches that attempt to figure out this relation in an automatic manner.
In particular, it solves the heterogeneity problem as a transfer learning \cite{pan2009survey} or multi-task \cite{caruana1997multitask} problem.\\
%
\subsubsection{Transfer learning}
The intuition behind this technique is that, even though the RSS readings from different devices may be different from each other, the learning tasks on these devices are related since the relation between the RSS coming from the different cell towers is the same for the different devices. Transfer learning allows one  to reuse a model pre-trained using a large number of samples coming from a device (master) as a starting point for fine-tuning the model for another device (slave). Unlike the traditional approximation techniques which require a large number of samples from the slave device for obtaining a good approximation, this technique requires only a few number of samples for tuning a new device.

Figure~\ref{fig:network-tl} presents the transfer learning model structure. We denote the RSS data collected by the master device $m$ as $D_m$ and denote RSS data collected by another device (slave) $s$ as $D_s$. $D_m$ is a large training data, while $D_s$ has only a few number of samples. 
 The network is trained using $D_m$. It attempts to learn and encode the relation between the input features into the latent-space (i.e. the bottleneck) $h$. 
The first layers of the network compress the input features into a latent-space representation. It can be represented by an encoding function $h=f(x)$.
Then, to fine-tune the model to a new device, one needs to tune the output layer (softmax layer) weights only using a few samples from the new device $D_s$ while freezing the weights of the remaining layers to get the estimated location.\\

\begin{figure}[!t]
\centering
\includegraphics[width=0.4\textwidth]{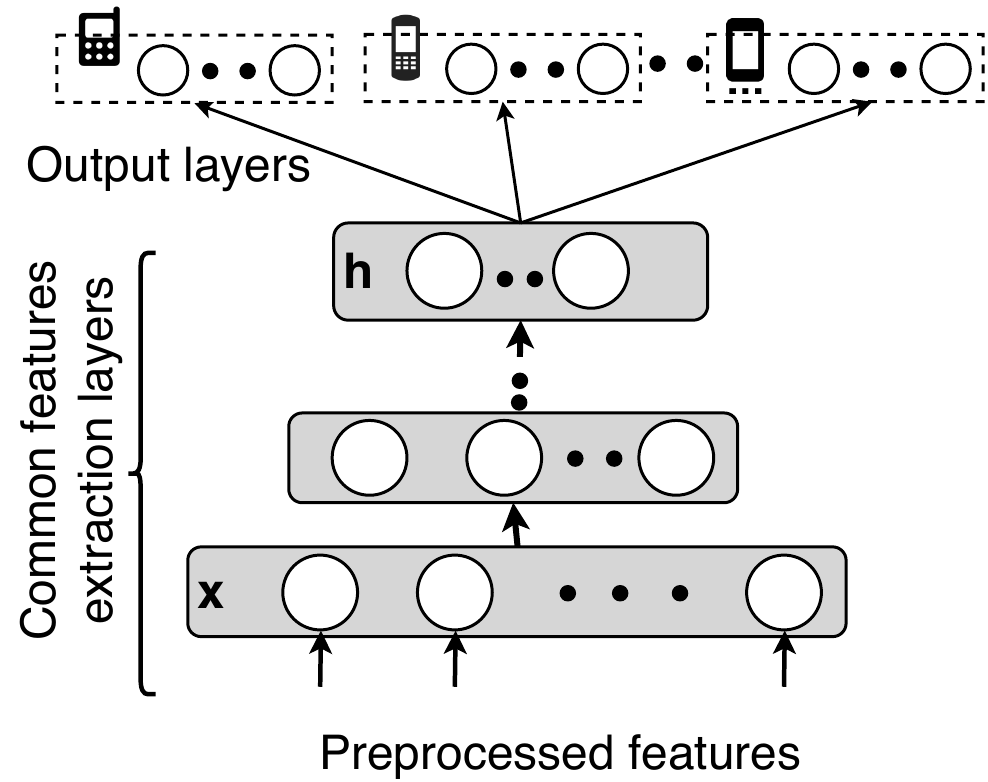}
     \caption{Multitask learning deep model structure. Input is the cell towers RSS information from different devices. Gray layers are responsible for extracting the relation between the RSS input features. The output layers (dashed layers) are trained simultaneously using a small number of samples from each phone type.}
     \label{fig:network-mtl}
\end{figure}
\subsubsection{Multitask learning}
This technique is a direct extension for the \textit{transfer learning} technique when the number of training samples from the slave device approaches number of samples from the master device ($|D_s| \simeq |D_m|$). It solves the heterogeneity problem as a multitask learning problem through learning the mapping between the RSS and locations from the different devices simultaneously. The model is trained with samples from different devices, which allows the model to be generalized for the heterogeneous devices. Figure~\ref{fig:network-mtl} presents the multitask learning model structure. The input to the model is the RSS from the different devices. There are different output layers for the different devices (tasks). All output layers share the same hidden layer, and hence they share the learning process.

\section{Performance evaluation} \label{evaluation}
In this section, we evaluate the performance of the proposed heterogeneity handling techniques. We start by describing the experimental setup used in the data collection process.
Then, we analyze the effect of different heterogeneity techniques on localization accuracy.

\subsection{Experimental setup}

Our testbeds cover 0.2$\text{Km}^2$ in a typical urban area and 1.2$\text{Km}^2$ in a typical rural area in Alexandria, Egypt. Cellular data and GPS ground-truth locations are collected by drivers.\\  
Data is collected by different devices. For space constrains, we present only results of four Android devices: HTC One X9 and Motorola Moto G5 Plus phones in the urban area, and HTC Nexus One and Prestigio Multipad Wize 3037 3G in the rural area. Number of samples collected by each device is around $10^5$ sample. We also collect a small dataset for the two devices with $10^3$ samples to approximate the linear function in linear transformation technique and fine-tunning the deep model in transfer learning technique.
The data collector App records the (cell-tower identifier, RSS, timestamp) for each heard tower in the area of interest. The scanning rate was set to one scan per second.
To discretize the area of interest, we used the gridding approach in \cite{ibrahim2012cellsense,deeploc} in which the area of interest is partitioned
virtually into a grid of equally-sized square cells and the center of each cell is considered as one reference location.

Without loss of generality, we employ the multinomial logistic classifier described in Section \ref{method} with three hidden layers as a deep model for localization. This model givens the best performance in \cite{deeploc}. The architecture parameters are shown in Table \ref{table:parameters}.

\begin{table}[!t]
            \centering
            \caption{Deep learning model parameters.}
            \scalebox{0.95}{
            \label{table:parameters}
            \begin{tabular}{|l|l|}
            \hline
             \textbf{Parameter}   & \textbf{Value} \\ \hline\hline
             \textit{ Learning rate }&  0.005 \\ \hline
             \textit{ Batch size }&  40 \\ \hline
             \textit{ Dropout rate (\%)}&  10 \\ \hline
             \textit{ Number of epochs }&  500 \\ \hline
             \textit{ Number of hidden layers }&  3 \\ \hline
             \textit{ Neurons per layer (urban) }& 25, 256, 128, 64, 20 \\ \hline
             \textit{ Neurons per layer (rural) }& 16, 256, 128, 64, 675 \\ \hline
             \textit{ Activation function }&  Sigmoid \\ \hline            
            \end{tabular}
            }
    \end{table}
    
\subsection{Localization accuracy}

\begin{figure*}[!t]
        \centering
        \begin{subfigure}[b]{0.4\textwidth}
            \centering
            \includegraphics[width=\textwidth]{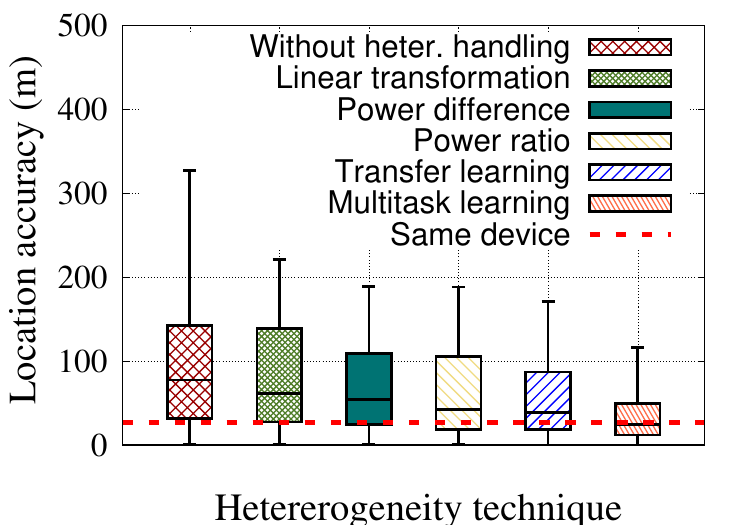}
            \caption[]%
            {{ Experiment I.}}    
            \label{exp1}
        \end{subfigure}
        \begin{subfigure}[b]{0.4\textwidth}  
            \centering 
            \includegraphics[width=\textwidth]{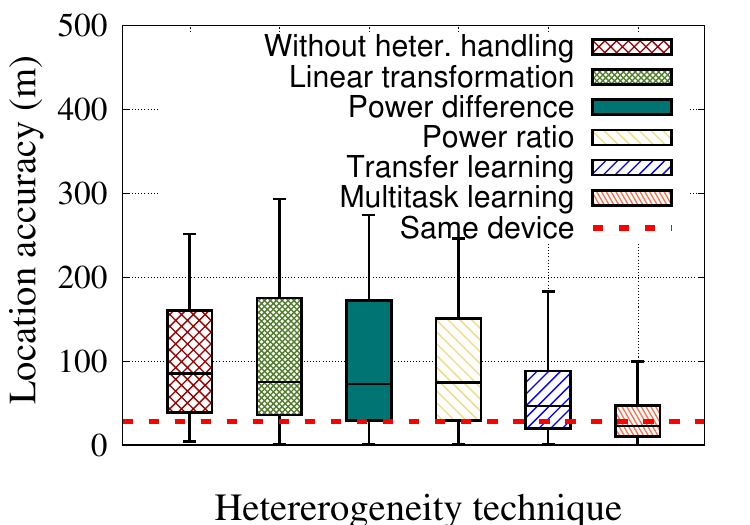}
            \caption[]%
            {{ Experiment II.}}    
            \label{exp2}
        \end{subfigure}
        
        \begin{subfigure}[b]{0.4\textwidth}   
            \centering 
            \includegraphics[width=\textwidth]{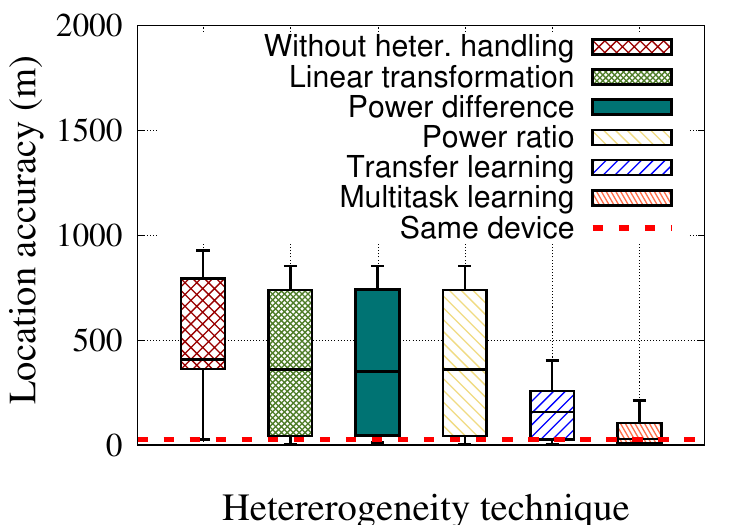}
            \caption[]%
            {{ Experiment III.}}    
            \label{exp3}
        \end{subfigure}
        \quad
        \begin{subfigure}[b]{0.4\textwidth}   
            \centering 
            \includegraphics[width=\textwidth]{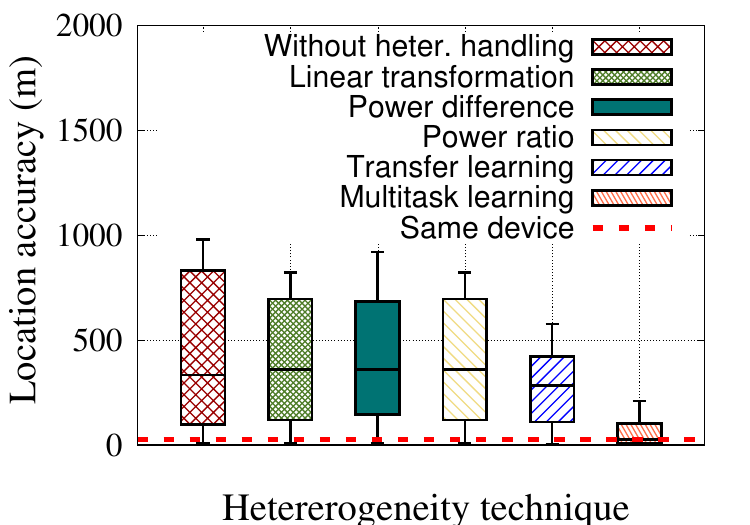}
            \caption[]%
            {{ Experiment IV.}}    
            \label{exp4}
        \end{subfigure}
        \caption[]
        { Effect of different device heterogeneity techniques on localization accuracy.} 
        \label{exp}
    \end{figure*}

We have performed four experiments: (I) training is performed using Motorola phone and testing is performed using HTC One X9 phone, (II) training is performed using HTC One X9 phone and testing is performed using Motorola phone, (III) training is performed using Prestigio Multipad Wize phone and testing is performed using HTC Nexus One phone,  and (IV) training is performed using HTC Nexus One phone and testing is performed using Prestigio Multipad Wize phone. Figure~\ref{exp} shows the effect of different device heterogeneity techniques on localization. accuracy.
Table~\ref{summary-results} summarizes the results. The results show that all heterogeneity-handling techniques improve the system accuracy compared to a system that does not handle the differences between devices.  
The results also highlight that the deep learning heterogeneity handling techniques are superior to other techniques. In particular, the Multitask Learning technique can give \textit{even better performance} than training and testing using the same device, with a median localization accuracy of 24m, 22m, 29m and 26m in the four experiments respectively. This is an enhancement of 220.8\%, 286.3\%, 1051\% and 1176\% in median error compared to training without handling heterogeneity. This enhancement in accuracy is due to the ability of transfer learning to automatically capture the mapping between the RSS of the same cell tower at different devices. In addition, the combined training between the two devices allow the multi-task learning model to handle noise and generalize in a better way.

\begin{table}[!t]
\centering
    \caption{Summary for localization results.}
    \scalebox{0.9}{
    \label{summary-results}
        \begin{tabular}{|l|l|l|l|l|}
            \hline
             Exp. & \pbox{10cm}{Heter\\handling } & \pbox{10cm}{$25^{th}$\\percentile(m) }& \pbox{10cm}{$50^{th}$\\percentile(m)}&\pbox{10cm}{$75^{th}$\\percentile(m)}\\ \hline
            \multirow{2}{*}{I} & \multicolumn{1}{l|}{Enabled} & \multicolumn{1}{l|}{\textbf{12}}  & \multicolumn{1}{l|}{\textbf{24}}& \multicolumn{1}{l|}{\textbf{50}}\\\cline{2-5}
                                 & \multicolumn{1}{l|}{Disabled} & \multicolumn{1}{l|}{31(-158.3\%)}  & \multicolumn{1}{l|}{77(-220.8\%)} & \multicolumn{1}{l|}{142(-184\%)}\\\hline
            \multirow{2}{*}{II} & \multicolumn{1}{l|}{Enabled.} & \multicolumn{1}{l|}{\textbf{10}}  & \multicolumn{1}{l|}{\textbf{22}}& \multicolumn{1}{l|}{\textbf{64}}\\\cline{2-5}
                                 & \multicolumn{1}{l|}{Disabled} & \multicolumn{1}{l|}{38(-280\%)}  & \multicolumn{1}{l|}{85(-286.3\%)} & \multicolumn{1}{l|}{161(-151.5\%)}\\\hline
                          
\multirow{2}{*}{III} & \multicolumn{1}{l|}{Enabled.} & \multicolumn{1}{l|}{\textbf{8}}  & \multicolumn{1}{l|}{\textbf{29}}& \multicolumn{1}{l|}{\textbf{104}}\\\cline{2-5}
                                 & \multicolumn{1}{l|}{Disabled} & \multicolumn{1}{l|}{43(-437\%)}  & \multicolumn{1}{l|}{334(-1051.3\%)} & \multicolumn{1}{l|}{739(-610.5\%)}\\\hline

                                 \multirow{2}{*}{IV} & \multicolumn{1}{l|}{Enabled.} & \multicolumn{1}{l|}{\textbf{10}}  & \multicolumn{1}{l|}{\textbf{26}}& \multicolumn{1}{l|}{\textbf{103}}\\\cline{2-5}
                                 & \multicolumn{1}{l|}{Disabled} & \multicolumn{1}{l|}{97(-870\%)}  & \multicolumn{1}{l|}{332(-1176\%)} & \multicolumn{1}{l|}{831(-706\%)}\\\hline
                                 
        \end{tabular}
        }
\end{table}

\section{Conclusion}
\label{conclusion}
In this paper, we have investigated different techniques for addressing the phones heterogeneity problem in deep cellular localization systems. The proposed solutions depend on either approximating a function that maps features from device to another or sharing knowledge across different devices. 
We have implemented and evaluated the impact of each device heterogeneity-handling technique on the deep model's localization accuracy. Results from four independent testbeds show that sharing the knowledge across different phones can improve the localization accuracy by more than 220\%. 
This highlights the use of transfer learning for solving the phones heterogeneity problem in deep cellular localization systems.

\bibliographystyle{ACM-Reference-Format}
\bibliography{main}


\begin{thebibliography}{18}


\ifx \showCODEN    \undefined \def \showCODEN     #1{\unskip}     \fi
\ifx \showDOI      \undefined \def \showDOI       #1{#1}\fi
\ifx \showISBNx    \undefined \def \showISBNx     #1{\unskip}     \fi
\ifx \showISBNxiii \undefined \def \showISBNxiii  #1{\unskip}     \fi
\ifx \showISSN     \undefined \def \showISSN      #1{\unskip}     \fi
\ifx \showLCCN     \undefined \def \showLCCN      #1{\unskip}     \fi
\ifx \shownote     \undefined \def \shownote      #1{#1}          \fi
\ifx \showarticletitle \undefined \def \showarticletitle #1{#1}   \fi
\ifx \showURL      \undefined \def \showURL       {\relax}        \fi
\providecommand\bibfield[2]{#2}
\providecommand\bibinfo[2]{#2}
\providecommand\natexlab[1]{#1}
\providecommand\showeprint[2][]{arXiv:#2}

\bibitem[\protect\citeauthoryear{Caruana}{Caruana}{1997}]%
        {caruana1997multitask}
\bibfield{author}{\bibinfo{person}{Rich Caruana}.}
  \bibinfo{year}{1997}\natexlab{}.
\newblock \showarticletitle{Multitask learning}.
\newblock \bibinfo{journal}{\emph{Machine learning}} \bibinfo{volume}{28},
  \bibinfo{number}{1} (\bibinfo{year}{1997}), \bibinfo{pages}{41--75}.
\newblock


\bibitem[\protect\citeauthoryear{Della~Rosa, Lepp{\"a}koski, Biancullo, and
  Nurmi}{Della~Rosa et~al\mbox{.}}{2010}]%
        {della2010ad}
\bibfield{author}{\bibinfo{person}{Francescantonio Della~Rosa},
  \bibinfo{person}{Helena Lepp{\"a}koski}, \bibinfo{person}{Stefano Biancullo},
  {and} \bibinfo{person}{Jari Nurmi}.} \bibinfo{year}{2010}\natexlab{}.
\newblock \showarticletitle{Ad-hoc networks aiding indoor calibrations of
  heterogeneous devices for fingerprinting applications}. In
  \bibinfo{booktitle}{\emph{2010 International Conference on Indoor Positioning
  and Indoor Navigation}}. IEEE, \bibinfo{pages}{1--6}.
\newblock


\bibitem[\protect\citeauthoryear{Figuera, Rojo-{\'A}lvarez, Mora-Jim{\'e}nez,
  Guerrero-Curieses, Wilby, and Ramos-L{\'o}pez}{Figuera et~al\mbox{.}}{2011}]%
        {figuera2011time}
\bibfield{author}{\bibinfo{person}{Carlos Figuera},
  \bibinfo{person}{Jos{\'e}~Luis Rojo-{\'A}lvarez}, \bibinfo{person}{Inmaculada
  Mora-Jim{\'e}nez}, \bibinfo{person}{Alicia Guerrero-Curieses},
  \bibinfo{person}{Mark Wilby}, {and} \bibinfo{person}{Javier
  Ramos-L{\'o}pez}.} \bibinfo{year}{2011}\natexlab{}.
\newblock \showarticletitle{Time-space sampling and mobile device calibration
  for WiFi indoor location systems}.
\newblock \bibinfo{journal}{\emph{IEEE Transactions on Mobile Computing}}
  \bibinfo{volume}{10}, \bibinfo{number}{7} (\bibinfo{year}{2011}),
  \bibinfo{pages}{913--926}.
\newblock


\bibitem[\protect\citeauthoryear{Gu, Shokry, and Youssef}{Gu
  et~al\mbox{.}}{2021}]%
        {gu2021effect}
\bibfield{author}{\bibinfo{person}{Chen Gu}, \bibinfo{person}{Ahmed Shokry},
  {and} \bibinfo{person}{Moustafa Youssef}.} \bibinfo{year}{2021}\natexlab{}.
\newblock \showarticletitle{The Effect of Ground Truth Accuracy on the
  Evaluation of Localization Systems}. In \bibinfo{booktitle}{\emph{IEEE
  INFOCOM 2021.}} IEEE, \bibinfo{pages}{1--10}.
\newblock


\bibitem[\protect\citeauthoryear{Ibrahim and Youssef}{Ibrahim and
  Youssef}{2012}]%
        {ibrahim2012cellsense}
\bibfield{author}{\bibinfo{person}{Mohamed Ibrahim} {and}
  \bibinfo{person}{Moustafa Youssef}.} \bibinfo{year}{2012}\natexlab{}.
\newblock \showarticletitle{CellSense: An accurate energy-efficient GSM
  positioning system}.
\newblock \bibinfo{journal}{\emph{IEEE Transactions on Vehicular Technology}}
  \bibinfo{volume}{61}, \bibinfo{number}{1} (\bibinfo{year}{2012}),
  \bibinfo{pages}{286--296}.
\newblock


\bibitem[\protect\citeauthoryear{Ibrahim and Youssef}{Ibrahim and
  Youssef}{2013}]%
        {ibrahim2013enabling}
\bibfield{author}{\bibinfo{person}{Mohamed Ibrahim} {and}
  \bibinfo{person}{Moustafa Youssef}.} \bibinfo{year}{2013}\natexlab{}.
\newblock \showarticletitle{Enabling wide deployment of GSM localization over
  heterogeneous phones}. In \bibinfo{booktitle}{\emph{Communications (ICC),
  2013 IEEE International Conference on}}. IEEE, \bibinfo{pages}{6396--6400}.
\newblock


\bibitem[\protect\citeauthoryear{Pan and Yang}{Pan and Yang}{2009}]%
        {pan2009survey}
\bibfield{author}{\bibinfo{person}{Sinno~Jialin Pan} {and}
  \bibinfo{person}{Qiang Yang}.} \bibinfo{year}{2009}\natexlab{}.
\newblock \showarticletitle{A survey on transfer learning}.
\newblock \bibinfo{journal}{\emph{IEEE Transactions on knowledge and data
  engineering}} \bibinfo{volume}{22}, \bibinfo{number}{10}
  (\bibinfo{year}{2009}).
\newblock


\bibitem[\protect\citeauthoryear{Park, Curtis, Teller, and Ledlie}{Park
  et~al\mbox{.}}{2011}]%
        {park2011implications}
\bibfield{author}{\bibinfo{person}{Jun-geun Park}, \bibinfo{person}{Dorothy
  Curtis}, \bibinfo{person}{Seth Teller}, {and} \bibinfo{person}{Jonathan
  Ledlie}.} \bibinfo{year}{2011}\natexlab{}.
\newblock \showarticletitle{Implications of device diversity for organic
  localization}. In \bibinfo{booktitle}{\emph{INFOCOM, 2011 Proceedings IEEE}}.
  IEEE, \bibinfo{pages}{3182--3190}.
\newblock


\bibitem[\protect\citeauthoryear{Shokry, Elhamshary, and Youssef}{Shokry
  et~al\mbox{.}}{2017}]%
        {shokry2017tale}
\bibfield{author}{\bibinfo{person}{Ahmed Shokry}, \bibinfo{person}{Moustafa
  Elhamshary}, {and} \bibinfo{person}{Moustafa Youssef}.}
  \bibinfo{year}{2017}\natexlab{}.
\newblock \showarticletitle{The Tale of Two Localization Technologies: Enabling
  Accurate Low-Overhead WiFi-based Localization for Low-end Phones}.
\newblock  (\bibinfo{year}{2017}).
\newblock


\bibitem[\protect\citeauthoryear{Shokry, Elhamshary, and Youssef}{Shokry
  et~al\mbox{.}}{2020}]%
        {shokry2020dynamicslam}
\bibfield{author}{\bibinfo{person}{Ahmed Shokry},
  \bibinfo{person}{Mostafa~Mahmoud Elhamshary}, {and} \bibinfo{person}{Moustafa
  Youssef}.} \bibinfo{year}{2020}\natexlab{}.
\newblock \showarticletitle{DynamicSLAM: Leveraging Human Anchors for
  Ubiquitous Low-Overhead Indoor Localization}.
\newblock \bibinfo{journal}{\emph{IEEE Transactions on Mobile Computing}}
  (\bibinfo{year}{2020}).
\newblock


\bibitem[\protect\citeauthoryear{Shokry, Torki, and Youssef}{Shokry
  et~al\mbox{.}}{2018}]%
        {deeploc}
\bibfield{author}{\bibinfo{person}{Ahmed Shokry}, \bibinfo{person}{Marwan
  Torki}, {and} \bibinfo{person}{Moustafa Youssef}.}
  \bibinfo{year}{2018}\natexlab{}.
\newblock \showarticletitle{DeepLoc: A Ubiquitous Accurate and Low-Overhead
  Outdoor Cellular Localization System.}. In \bibinfo{booktitle}{\emph{ACM
  SIGSPATIAL.}}
\newblock


\bibitem[\protect\citeauthoryear{Shokry and Youssef}{Shokry and
  Youssef}{2021}]%
        {quantum_arx}
\bibfield{author}{\bibinfo{person}{Ahmed Shokry} {and}
  \bibinfo{person}{Moustafa Youssef}.} \bibinfo{year}{2021}\natexlab{}.
\newblock \showarticletitle{{Challenge: Quantum Computing for Location
  Determination}}.
\newblock \bibinfo{journal}{\emph{arXiv e-prints}} (\bibinfo{year}{2021}),
  \bibinfo{pages}{arXiv--2106}.
\newblock


\bibitem[\protect\citeauthoryear{Shokry and Youssef}{Shokry and
  Youssef}{2022a}]%
        {quantum_lcn}
\bibfield{author}{\bibinfo{person}{Ahmed Shokry} {and}
  \bibinfo{person}{Moustafa Youssef}.} \bibinfo{year}{2022}\natexlab{a}.
\newblock \showarticletitle{{A Quantum Algorithm for RF-based Fingerprinting
  Localization Systems}}.
\newblock \bibinfo{journal}{\emph{IEEE Conference on Local Computer Networks
  (LCN)}} (\bibinfo{year}{2022}).
\newblock


\bibitem[\protect\citeauthoryear{Shokry and Youssef}{Shokry and
  Youssef}{2022b}]%
        {device_indp_q}
\bibfield{author}{\bibinfo{person}{Ahmed Shokry} {and}
  \bibinfo{person}{Moustafa Youssef}.} \bibinfo{year}{2022}\natexlab{b}.
\newblock \showarticletitle{{Device-independent Quantum Fingerprinting for
  Large Scale Localization}}. In \bibinfo{booktitle}{\emph{2022 MedComNet}}.
  IEEE, \bibinfo{pages}{208--215}.
\newblock


\bibitem[\protect\citeauthoryear{Shokry and Youssef}{Shokry and
  Youssef}{2022c}]%
        {quantum_qce}
\bibfield{author}{\bibinfo{person}{Ahmed Shokry} {and}
  \bibinfo{person}{Moustafa Youssef}.} \bibinfo{year}{2022}\natexlab{c}.
\newblock \showarticletitle{{QLoc: A Realistic Quantum Fingerprint-based
  Algorithm for Large Scale Localization}}.
\newblock \bibinfo{journal}{\emph{IEEE QCE}} (\bibinfo{year}{2022}).
\newblock


\bibitem[\protect\citeauthoryear{Shokry and Youssef}{Shokry and
  Youssef}{2023}]%
        {qhet}
\bibfield{author}{\bibinfo{person}{Ahmed Shokry} {and}
  \bibinfo{person}{Moustafa Youssef}.} \bibinfo{year}{2023}\natexlab{}.
\newblock \showarticletitle{Quantum fingerprinting for heterogeneous devices
  localization}.
\newblock \bibinfo{journal}{\emph{Computer Communications}}
  (\bibinfo{year}{2023}).
\newblock
\showISSN{0140-3664}
\urldef\tempurl%
\url{https://doi.org/10.1016/j.comcom.2023.03.010}
\showDOI{\tempurl}


\bibitem[\protect\citeauthoryear{Tsui, Chuang, and Chu}{Tsui
  et~al\mbox{.}}{2009}]%
        {tsui2009unsupervised}
\bibfield{author}{\bibinfo{person}{Arvin~Wen Tsui}, \bibinfo{person}{Yu-Hsiang
  Chuang}, {and} \bibinfo{person}{Hao-Hua Chu}.}
  \bibinfo{year}{2009}\natexlab{}.
\newblock \showarticletitle{Unsupervised learning for solving RSS hardware
  variance problem in WiFi localization}.
\newblock \bibinfo{journal}{\emph{Mobile Networks and Applications}}
  \bibinfo{volume}{14}, \bibinfo{number}{5} (\bibinfo{year}{2009}),
  \bibinfo{pages}{677--691}.
\newblock


\bibitem[\protect\citeauthoryear{Wang, Wang, and Mao}{Wang
  et~al\mbox{.}}{2017}]%
        {wang2017resloc}
\bibfield{author}{\bibinfo{person}{Xuyu Wang}, \bibinfo{person}{Xiangyu Wang},
  {and} \bibinfo{person}{Shiwen Mao}.} \bibinfo{year}{2017}\natexlab{}.
\newblock \showarticletitle{ResLoc: Deep residual sharing learning for indoor
  localization with CSI tensors}. In \bibinfo{booktitle}{\emph{Personal,
  Indoor, and Mobile Radio Communications (PIMRC), 2017 IEEE 28th Annual
  International Symposium on}}. IEEE, \bibinfo{pages}{1--6}.
\newblock


\end{thebibliography}

\end{document}